\documentclass[aps,prl,reprint,superscriptaddress]{revtex4-1}
\pdfoutput=1
\usepackage{easy-todo}
\usepackage{color}
\usepackage{graphicx}
\usepackage{amsmath}
\usepackage{amsfonts}
\usepackage{amssymb}
\usepackage{microtype}
\usepackage{verbatim}
\usepackage{bbold}
\usepackage{hyperref}
\usepackage{cleveref}
\allowdisplaybreaks
\usepackage{multirow}
\usepackage{bbm}
\usepackage{braket}
\hypersetup{
  colorlinks   = true, 
  urlcolor     = blue, 
  linkcolor    = blue, 
  citecolor   = blue 
}

\DeclareMathOperator{\E}{\mathbb{E}}

\begin{document}

\author{Pieter W. Claeys}
\email{pc652@cam.ac.uk}
\affiliation{TCM Group, Cavendish Laboratory, University of Cambridge, Cambridge CB3 0HE, UK}

\author{Austen Lamacraft}
\affiliation{TCM Group, Cavendish Laboratory, University of Cambridge, Cambridge CB3 0HE, UK}

\author{Jonah~Herzog-Arbeitman}
\affiliation{TCM Group, Cavendish Laboratory, University of Cambridge, Cambridge CB3 0HE, UK}

\title{Absence of superdiffusion in certain random spin models}

\begin{abstract}
The dynamics of spin at finite temperature in the spin-1/2 Heisenberg chain was found to be \emph{superdiffusive} in numerous recent numerical and experimental studies. Theoretical approaches to this problem have emphasized the role of nonabelian $SU(2)$ symmetry as well as integrability but the associated methods cannot be readily applied when integrability is broken.

We examine spin transport in a spin-1/2 chain in which the exchange couplings fluctuate in space and time around a nonzero mean $J$, a model introduced by De Nardis \emph{et al.} \cite{de_nardis_stability_2021}. We show that operator dynamics in the strong noise limit at infinite temperature can be analyzed using conventional perturbation theory as an expansion in $J$. We find that regular diffusion persists at long times, albeit with an enhanced diffusion constant. The finite time spin dynamics is analyzed and compared with matrix product operator simulations. 

\end{abstract}

\maketitle
\emph{Introduction.} The large-scale dynamics of a system close to equilibrium can usually be understood in terms of a few simple types of motion, including diffusion and sound waves, with the number and nature of these modes determined by conservation laws. In the simplest case of a single conservation law, and ignoring the possibility of broken symmetry by working at sufficiently high temperatures, diffusion of the conserved quantity is the norm. This simple picture applies even in the case of a multicomponent conserved quantity such as the spin density in an isotropic paramagnet: within linear response each component diffuses separately \cite{forster2018hydrodynamic}. 

Recently, an exception to this phenomenology has generated a great deal of interest. For the one dimensional spin-1/2 Heisenberg model, numerical calculations revealed that infinite temperature spin dynamics is \emph{superdiffusive}, with dynamic scaling $t\sim \ell^{z}$ of time ($t$) and length ($\ell$) where $z=3/2$ (c.f. $z=2$ for ordinary diffusion) \cite{PhysRevLett.106.220601,ljubotina2017spin}, characteristic of the Kardar--Parisi--Zhang (KPZ) universality class \cite{kardar1986dynamic}. Detailed numerical comparisons showed agreement between the spin correlation function and the known exact scaling function describing stationary correlations of the KPZ equation \cite{ljubotina_kardar-parisi-zhang_2019}.
 
The same phenomenology was found to be consistent with numerically evaluated spin correlations in classical integrable models of spin dynamics \cite{das2019kardar,krajnik2020kardar} and integrable models with larger symmetry groups \cite{krajnik_integrable_2020,PhysRevB.101.121106}, as well as two recent experiments \cite{scheie2021detection,wei2021quantum}. This led to an emerging consensus that superdiffusion arises in one dimension due to \emph{a combination of nonabelian symmetry and integrability}. A theoretical understanding of superdiffusion in such models has been achieved within the assumptions of (generalized) hydrodynamics \cite{ilievski_superuniversality_2020,PhysRevLett.121.230602,PhysRevLett.122.127202}. A fully microscopic calculation of the KPZ scaling function for any model is still lacking, however. A recent review summarizes the state of the art \cite{Bulchandani_2021}.

While the effect of breaking integrability was investigated numerically in several earlier works \cite{krajnik2020kardar,PhysRevB.101.121106}, a theoretical analysis appeared only recently. In Ref.~\cite{de_nardis_stability_2021} the effect of random perturbations in space and time was studied. For a weak noisy exchange coupling that preserves spin rotation symmetry the diffusion constant was found to grow logarithmically in time $D(t)\sim \log t$, indicating superdiffusion, albeit of a weaker variety. From a hydrodynamic perspective, this model should be described by (interacting, nonlinear) diffusing spin modes, as there are no other conserved quantities, such as energy. This makes the noisy exchange model an ideal testing ground for the validity of the hydrodynamic theory. In Ref.~\cite{glorioso2021hydrodynamics} it was argued that such a description leads to \emph{normal} diffusion, with subleading corrections $\propto 1/\sqrt{t}$.

In this Letter, we settle this issue by providing a detailed microscopic theory of spin dynamics in the noisy exchange model. We show that operator dynamics near the strong noise limit at infinite temperature can be analyzed using perturbation theory as an expansion in $J$, the constant part of the exchange coupling. We find that regular diffusion persists at long times, albeit with an enhanced diffusion constant, and that other features of the spin dynamics are consistent with the hydrodynamic theory of Ref.~\cite{glorioso2021hydrodynamics}. These results are compared with matrix product operator simulations, showing excellent agreement, including the interpolation from the bare to renormalized diffusion constant as time progresses.

\emph{Correlation functions.} We consider a spin-1/2 chain of $N$ sites with spin $\boldsymbol{\sigma}_n=(X_n, Y_n, Z_n)$ at site $n$. We are interested in the infinite temperature spin-spin correlator
\begin{equation}\label{eq:corr-def}
C^{ab}_{mn}(t)\equiv\frac{1}{2^N}\mathop{\mathrm{tr}}\left[\sigma^a_m(0) \sigma^b_n(t)\right].
\end{equation}
$\sigma^b_n(t)=\mathcal{U}^\dagger_t \sigma^b_n \mathcal{U}_t$ denotes the time dependence of the spin operator at site $n$ in the Heisenberg picture. $SU(2)$ invariance implies $C^{ab}_{mn}(t)\equiv\delta_{ab}C_{mn}(t)$ with $\sum_{n=1}^N C_{mn}(t)=1$. From now on we fix $a=b=z$ for definiteness.

The operator $Z_n(t)$ can be expanded in a basis of products of local operators as
$$
Z_n(t)= \sum_{\mu_{1:N}\in\{0,1,2,3\}^N} \mathcal{C}_{\mu_{1:N}}(t) \sigma_1^{\mu_1}\otimes\cdots \sigma_N^{\mu_N},
$$
where $\sigma^\mu = (\mathbbm{1},X,Y,Z)$. With the initial condition 
\begin{equation}\label{eq:initial}
\mathcal{C}_{\mu_{1:N}}(0)=\begin{cases}
1 & \mu_n=z,\, \mu_m=0,\,\forall m\neq n, \\
0 & \text{otherwise},
\end{cases} 
\end{equation}
the spin correlation function is
$$
C_{mn}(t) = \mathcal{C}_{0\cdots \mu_m=z \cdots 0}(t).
$$

\emph{Model.} A spin chain with fluctuating exchange coupling is described the stochastic Schr\"odinger equation \cite{breuer_theory_2002} (setting $\hbar =1$) 
\begin{equation}\label{eq:sse}
d\ket{\psi} = \sum_n \left[-i(J dt + \sqrt{\eta}dW_n)P_{n,n+1}-\frac{\eta}{2}dt\right]\ket{\psi}.
\end{equation}
Here $P_{n,n+1}=\left[1+\sum_a \sigma^a_n \sigma^a_{n+1}\right]/2$ is the exchange operator, and $W_n$ are independent Brownian motions with strength $\mu$ that give rise to white noise fluctuations $\propto dW_n$ of the exchange between sites $n$ and $n+1$. The last term is required to preserve $\braket{\psi|\psi}$ when Eq.~\eqref{eq:sse} is interpreted as an It\^o stochastic differential equation. From now on we measure time in units such that $\eta=1$.

The corresponding Heisenberg equation of motion is
\begin{multline}\label{eq:hberg}
d\mathcal{O} = \sum_n \Big[i\left(J dt + dW_n\right)\left[P_{n,n+1},\mathcal{O}\right]\\+dt\left(P_{n,n+1}\mathcal{O}P_{n,n+1}-\mathcal{O}\right)\Big].
\end{multline}
The average $\bar{\mathcal{O}}\equiv\E \mathcal{O}$ obeys the (adjoint of the) Lindblad equation
\begin{equation}
\frac{d\bar{\mathcal{O}}}{dt} = \sum_n \Big[iJ \left[P_{n,n+1},\bar{\mathcal{O}}\right]+\left(P_{n,n+1}\bar{\mathcal{O}}P_{n,n+1}-\bar{\mathcal{O}}\right)\Big].
\label{eq:eom}
\end{equation}
Alternatively, we derive this equation via the continuous time limit of a random circuit model in the Supplemental Material \cite{supplemental}, where the unitary circuits fluctuate around circuits arising from a Trotterization of the Heisenberg Hamiltonian \cite{vanicat_integrable_2018,ljubotina_ballistic_2019,friedman_spectral_2019,krajnik_integrable_2020,claeys_correlations_2021}. In Eq.~\eqref{eq:eom} the term $P_{n,n+1}\bar{\mathcal{O}}P_{n,n+1}$ exchanges the operators on sites $n$ and $n+1$. For $J=0$ the result is a master equation describing random adjacent transpositions, and preserves subspaces corresponding to fixed numbers of each of the $\sigma^\mu$. Starting from the initial condition Eq.~\eqref{eq:initial} and introducing the shorthand notation $\mathcal{C}_{0\cdots \mu_m=a\cdots 0}\equiv C^a_m$ we have the equation of motion
$$
\partial_t C^a_m = C^a_{m+1} + C^a_{m-1} - 2 C^a_m\equiv \Delta_m C^a_m 
$$
describing diffusion of a single $\sigma^a$, where we have defined the 1D discrete Laplacian $\Delta_m$

\begin{figure}[t]
  \includegraphics[width=0.9\columnwidth]{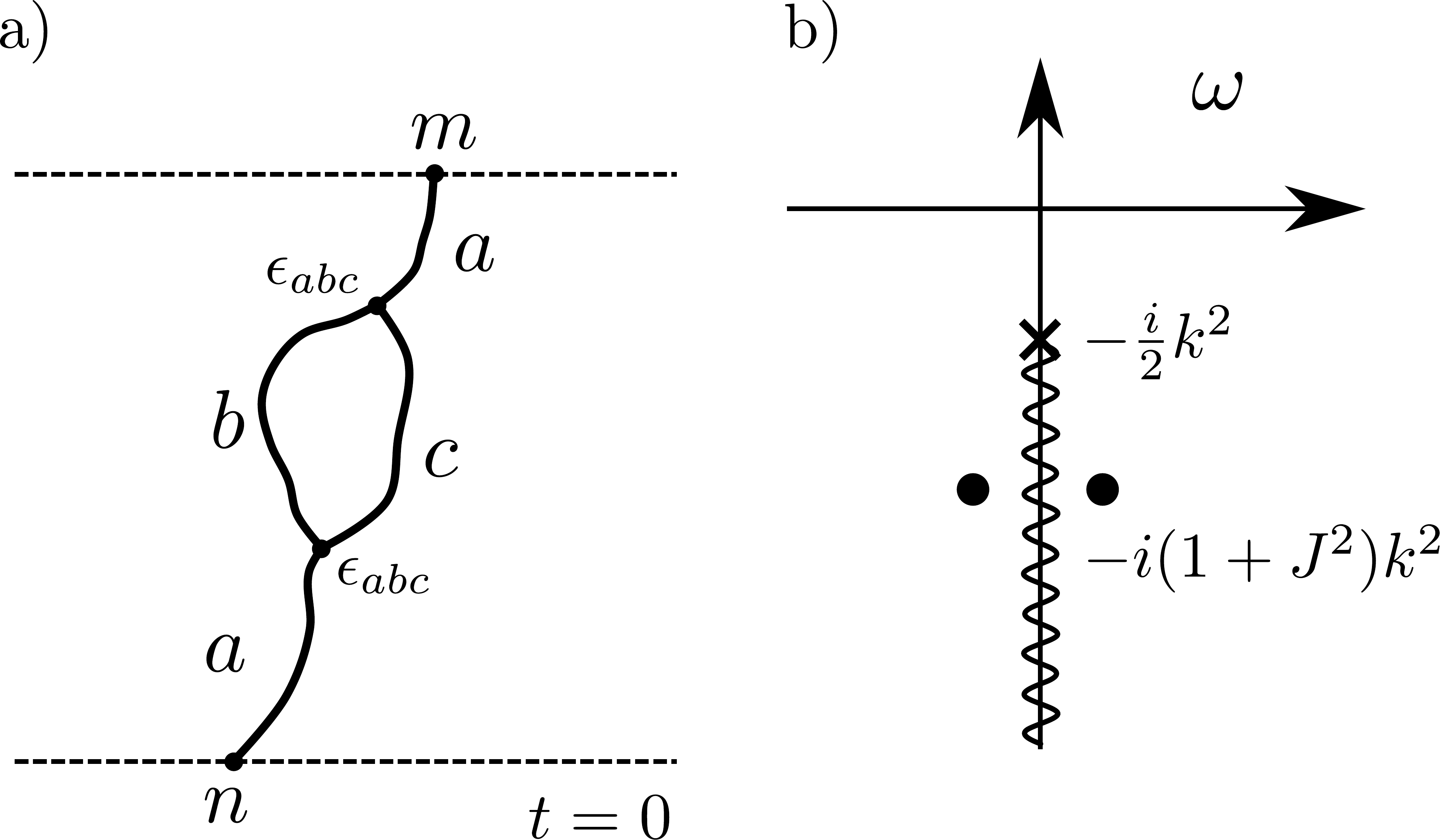}
  \caption{(a) Operator dynamics described by Eq.~\eqref{eq:eom}. (b) Analytic structure of the one-loop self energy Eq.~\eqref{eq:se-final-low}.  
  \label{fig:operator-dynamics}}
  \end{figure}

\emph{Perturbation theory.} To evaluate the effect of $J\neq 0$ we use
$$
i\left[P_{n,n+1},\sigma^\mu_{n}\otimes \sigma^\nu_{n+1}\right]=\epsilon_{\mu\nu\lambda\kappa}\sigma_{n}^\lambda\otimes\sigma_{n+1}^\kappa \ .
$$
The component form of Eq.~\eqref{eq:eom} is then
\begin{multline}
\partial_t \mathcal{C}_{\mu_{1:N}} = \sum_n \left[J\epsilon_{\alpha\beta \mu_n \mu_{n+1}} \mathcal{C}_{\mu_1\cdots \alpha\beta \cdots \mu_N} \right.\\\left.+ \mathcal{C}_{\mu_1\cdots \mu_{n+1}\mu_n \cdots \mu_N} - \mathcal{C}_{\mu_1\cdots \mu_{n}\mu_{n+1} \cdots \mu_N}\right].
\label{eq:component-eom}
\end{multline}
Thus a single $Z$ can give rise to an $X$, $Y$ pair. In general, the first term of Eq.~\eqref{eq:component-eom} describes `split' and `merge' processes with the spin structure of each vertex being given by $\epsilon_{abc}$ (\Cref{fig:operator-dynamics}). These processes couple sectors with different numbers of non-identity operators. We now develop a simple approximation, valid at small $J$, that keeps track of the one- and two-operator sectors, described by $C^{bc}_{m,n}\equiv \mathcal{C}_{0\cdots \mu_m=b\cdots \mu_n=c\cdots 0}$ by writing
\begin{align}
  &\partial_t C^z_n = J\left[C^{xy}_{n-1,n}-C^{xy}_{n,n-1}-C^{xy}_{n,n+1}+C^{xy}_{n+1,n}\right] \nonumber\\
   &\qquad\qquad+\Delta_n C^z_n\,,
  \label{eq:1op} \\
  &\partial_t C^{xy}_{m,n} = J \left[\delta_{m+1,n}\left(C^z_m-C^z_{m+1}\right)+\delta_{m,n+1}\left(C^z_{n+1}-C^z_n\right)\right] \nonumber\\
  &\,\,\,\,\,+\Delta_m C^{xy}_{m,n}+ \Delta_n C^{xy}_{m,n} + \delta_{|m-n|,1}(C^{xy}_{m,n}+C^{xy}_{n,m})\,.
  \label{eq:2op}
  \end{align}
  In addition, Eq.~\eqref{eq:2op} has the boundary condition $C^{xy}_{m,m}=0$ that, together with the last term, arises from the impossibility of having two operators on the same site. Because of the antisymmetry of the vertex [c.f. Eq.~\eqref{eq:1op}], however, this constraint plays no role in the solution of Eqs.~\eqref{eq:1op} and \eqref{eq:2op} \cite{supplemental}. The result is that $C^z(\omega,k)$, the Fourier transform of the correlation function in space and time, can be written in terms of a self-energy $\Sigma(\omega,k)$ as $C^z(\omega,k)=\left[i\omega - \Omega(k)-\Sigma(\omega,k)\right]^{-1}$ with $\Omega(k)\equiv 4\sin^2(k/2)$ and 
\begin{equation}
  \Sigma(\omega,k) =  \frac{4J^2}{N} \sum_{q} \frac{\left[\cos(q)-\cos(k-q)\right]^2}{\Omega(q)+\Omega(k-q)-i\omega}\,,
  \label{eq:se}
\end{equation}
where $q=  \frac{2\pi n}{N}$ is a momentum in the Brillouin zone.
For $N\to\infty$ and in the low momentum, low energy limit where $\Omega(k)\to k^2$ and $\omega$ is $O(k^2)$ we find \cite{supplemental}
\begin{equation}
  \Sigma(\omega, k) = J^2k^2\left[1+\frac{1}{2}\sqrt{k^2-2i\omega}\right].
  \label{eq:se-final-low}
\end{equation}
The analytic structure of $C^z(\omega,k)$ implied by Eq.~\eqref{eq:se-final-low} is as follows (\Cref{fig:operator-dynamics}):

\begin{enumerate}
  \item The diffusion pole located at $\omega=-ik^2$ when $J=0$ becomes a pair of poles at
  \begin{equation}
    \omega = -i(1+J^2)k^2 \pm |k|^3\frac{J^2}{2}\sqrt{1+2J^2} + O(k^4).
    \label{eq:poles}
  \end{equation}

  \item There is a branch cut starting at $\omega=-ik^2/2$, which is the threshold for a mode of wavevector $k$ to decay to two modes with wavevectors $k_{1,2}=k/2$.
\end{enumerate}

The same structure appears in Ref.~\cite{chen2019theory} in a calculation based on nonlinear hydrodynamic fluctuation theory in diffusive systems. Eq.~\eqref{eq:poles} shows that $J$ enhances ordinary diffusion, to be contrasted with the superdiffusive behavior of the diffusion constant $D(t)\sim \log t$ found at weak noise in Ref.~\cite{de_nardis_stability_2021}. To understand the analogous crossover in time from bare to enhanced diffusion, we evaluate the diffusion constant as $D(t) = -\frac{1}{2} \left. \partial_t\partial^2_k C^z(t,k) \right\vert_{k=0}$.  
As in regular diffusion, $D(t)$ determines the rate at which the width of the real-space profile of the correlation grows, since we also have
\begin{align}\label{eq:diffusionaswidth}
D(t) = \frac{1}{2}\frac{d}{dt}\left[\sum_n n^2 C_n^z(t) - \left(\sum_n n C_n^z(t)\right)^2\right]\,.
\end{align}
Using Eqs.~\eqref{eq:1op} and \eqref{eq:2op} the diffusion constant may be analytically calculated to order $J^2$ \cite{supplemental}:
\begin{align}\label{eq:diffusionconstant}
  D(t) &= 1 + J^2\big(1 - e^{-4t}\left[I_0(4t)+I_1(4t)\right] \big)\nonumber\\ &\quad \underset{t\to\infty}{\longrightarrow}  1 + J^2 - \frac{J^2}{\sqrt{2\pi t}}, 
  \end{align}
  expressed in terms of modified Bessel functions of the first kind and where the long-time limit follows from their asymptotic expression at large argument, $I_{\alpha}(z) \approx e^z / \sqrt{2\pi z}$. At $t=0$ we also recover $D(t=0) = 1$.

It's also useful to find the real space profile of the correlation function, assuming it is dominated by the poles in Eq.~\eqref{eq:poles}. Taking the continuum limit and evaluating the Fourier transform using a saddle-point approximation, we find that
\begin{align}\label{eq:saddlepointprofile}
  C^z(x;t) \propto \exp\left(-\frac{x^2}{2Dt}\right)\exp\left(-\frac{J^2 \sqrt{2J^2+1}}{2 D^3}\frac{|x|^3}{t^2}\right),
  \end{align}
expressed in terms of the asymptotic diffusion constant $D=1+J^2$. The second factor is strongly reminiscent of the $z=3/2$ dynamic scaling of the KPZ universality class, and for small $J$ this profile returns the previously obtained diffusion constant to good accuracy. Expanding all expressions in orders of $J^2$, evaluating Eq.~\eqref{eq:diffusionaswidth} using the correlation profile \eqref{eq:saddlepointprofile} returns
\begin{align}
D(t) = D - \frac{3}{2\sqrt{2}}\frac{J^2}{\sqrt{2\pi Dt}} \approx 1+J^2-1.06\frac{J^2}{\sqrt{2\pi t}}\,.
\end{align}

\emph{Numerics.} In order to test the validity of our perturbative approach we compare these predictions with matrix product operator (MPO) simulations of the operator dynamics, where we map the operator to a state in a doubled Hilbert space and numerically integrate the Lindblad equation~\eqref{eq:eom} using the time-evolution block-decimation (TEBD) method \cite{vidal_efficient_2004,verstraete_matrix_2004,zwolak_mixed-state_2004}. In all calculations the operator $Z_j(t)$ is represented as a matrix product state with maximal bond dimension $\chi = 400$ and a truncation error $\epsilon = 10^{-12}$, sufficient to ensure convergence for all presented results. The TEBD evolution uses a Suzuki-Trotter decomposition of the exact evolution operator with a discrete time step $\delta t = 10^{-2}$ and is implemented using a modified version of TeNPy \cite{hauschild_efficient_2018}. Note that for $J=0$ this approach is exact (up to a Trotterization error) since the operator can at all times be exactly represented as an MPO with maximal bond dimension $\chi = 2$. The Trotterization is not expected to qualitatively change the results: superdiffusion and KPZ scaling were also observed in integrable discrete time (circuit) models \cite{krajnik2020kardar}.

\begin{figure}[t]
  \includegraphics[width=\columnwidth]{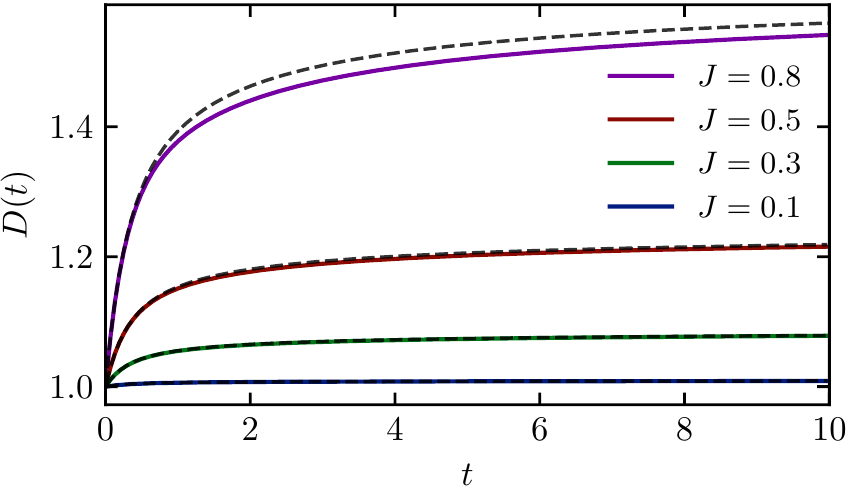}
  \caption{Time-dependent diffusion constant $D(t)$ for a system of $L=100$ spins at four different values of $J$. Full lines denote the MPO simulation, dashed lines the corresponding perturbative result~\eqref{eq:diffusionconstant}.
  \label{fig:diffusionconstant}}
  \end{figure}
  
We first compare the time-dependent diffusion constant $D(t)$ with the perturbative prediction~\eqref{eq:diffusionconstant} as a function of time and at different values of $J$ in Fig.~\ref{fig:diffusionconstant}. For $J$ sufficiently small the agreement is excellent at all times, providing an a posteriori justification of our perturbative approach. As expected, increasing $J$ leads to larger deviations of the MPO result from Eq.~\eqref{eq:diffusionconstant}. However, at all times the first-order calculation presents an upper bound to the exact diffusion constant, suggesting that no superdiffusion will appear on these numerically accessible time scales.

\begin{figure}[t]
  \includegraphics[width=\columnwidth]{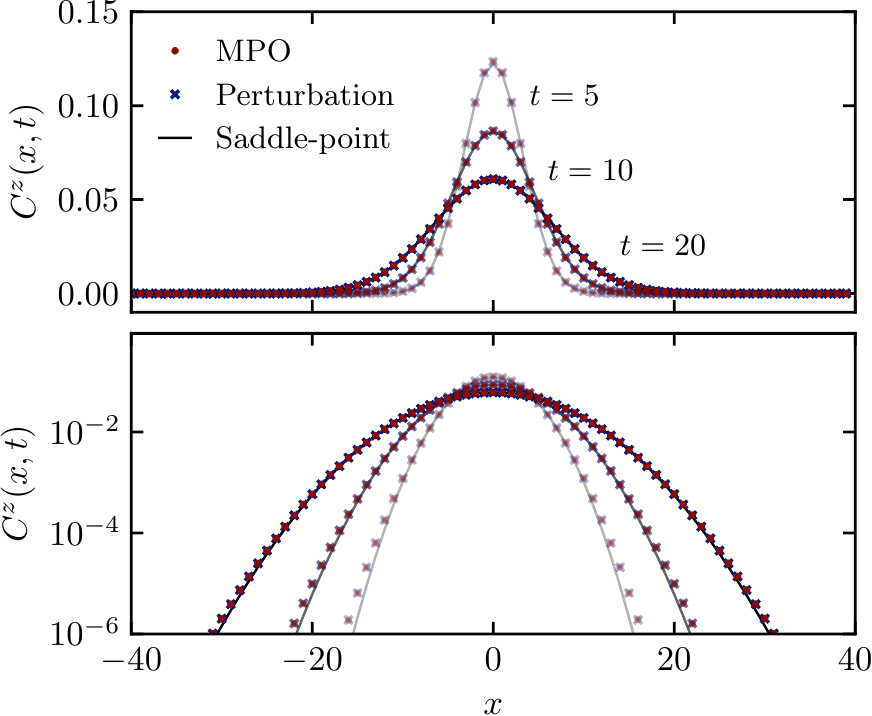}
  \caption{Real-space correlation profile $C^z(x,t) = C^z_{n=x}(t)$ using both a linear and a logarithmic scale for a system of $L=80$ spins with $J=0.3$. The three lines correspond to the profile evaluated at three different times $t=5,10,20$. Red dots denote the MPO simulation, blue crosses the numerical integration of the one- and two-component equations, and full black line Eq.~\eqref{eq:saddlepointprofile}.
  \label{fig:profile}}
  \end{figure}
  
We now move to the real-space profile of the correlation function. It is straightforward to numerically integrate the combined equations of motion~\eqref{eq:1op} and \eqref{eq:2op} to return the correlation profile within our perturbative approach, and in Fig.~\ref{fig:profile} we compare the resulting correlations $C^z_n$ for a fixed $J$ with both the MPO results and the profile \eqref{eq:saddlepointprofile} obtained using a saddle-point approximation. The MPO results for the full Lindblad equation clearly agree with the perturbative prediction over multiple orders of magnitude. The saddle-point profile then provides a good approximation to these results, with slight deviations at the tail ends of the distribution. However, note that this profile is not a fit since it has no free parameters, and the agreement increases with increasing time. Increasing $J$ again leads to an increasing deviation of the MPO results from the perturbative prediction.

\emph{Discussion.} We have shown that the fluctuating exchange model of Ref.~\cite{de_nardis_stability_2021} is amenable to theoretical analysis in the regime of strong noise (small $J$), and provided convincing analytical and numerical evidence that this regime is described by ordinary diffusion, albeit with some interesting transient behaviour.

It is possible to compare our microscopic approach, based on the equations of motion \eqref{eq:eom}, with the hydrodynamic theory of Ref.~\cite{glorioso2021hydrodynamics}. In that work, the long wavelength limit of the spin current $j^a$ is assumed to have an expansion
$$
j^a = -D\nabla s^a + \lambda \epsilon_{abc} s^b \nabla s^c + \cdots
$$
in terms of the spin density $s^a$. This form of the current is consistent with our Eq.~\eqref{eq:1op} and $\lambda=J$, by interpreting the latter as a continuity equation and identifying the right hand side as $-\nabla \cdot j^z$, together with the mean field assumption $C^{xy}_{n,n+1}\sim s^x s^y$. In Ref.~\cite{glorioso2021hydrodynamics} it is argued that nonintegrable spin chains have ordinary diffusive spin dynamics with a subleading correction $\propto \lambda^2/\sqrt{t}$ that is consistent with our Eq.~\eqref{eq:diffusionconstant}.

Our perturbative calculation includes the decay of one diffusive mode into two, leading to a branch cut starting at $\omega=-ik^2/2$. The true low frequency response of the system will be determined by decay into many modes, with the $n$-diffuson cut starting at $\omega=-ik^2/n$. On the basis of general estimates of these processes, Ref.~\cite{delacretaz2020heavy} concludes that the long time relaxation of a mode of wavevector $k$ is described by $\sim\exp(-\text{const.} \times \sqrt{Dk^2t})$. It would be interesting to establish whether this is so in any microscopic model.

\emph{Acknowledgements.} -- We gratefully acknowledge support from EPSRC Grant No. EP/P034616/1. JHA is supported by a Marshall Scholarship. This research was supported in part by the National Science Foundation under Grant No. NSF PHY-1748958. AL thanks Sarang Gopalakrishnan for numerous enlightening discussions.

\bibliography{Library.bib}

\newpage
\begin{widetext}
\section{Supplemental Material}

\subsection{Random circuit model}
The Lindblad equation \eqref{eq:eom} can alternatively be obtained from unitary circuit dynamics, where the dynamics is generated by two-site unitary gates acting on neighboring sites $j,j+1$ as
\begin{align}
U_{j,j+1} = \cos(\theta) \mathbbm{1}_{j,j+1} - i \sin(\theta) P_{j,j+1},
\end{align}
with $P_{j,j+1}$ the permutation operator. These operators preserve $SU(2)$ symmetry and arise naturally in the Trotterization of the integrable Heisenberg XXX chain \cite{vanicat_integrable_2018,ljubotina_ballistic_2019,friedman_spectral_2019,krajnik_integrable_2020,claeys_correlations_2021}. The resulting operator dynamics follows as
\begin{align}
\mathcal{O} \to  U_{j,j+1}^{\dagger} \mathcal{O} U_{j,j+1} &= \cos^2(\theta) \mathcal{O} + \sin^2(\theta) P_{j,j+1} \mathcal{O}P_{j,j+1} + i \sin(\theta)\cos(\theta) [P_{j,j+1},\mathcal{O}] \\
&=\mathcal{O} + \frac{i}{2} \sin(2\theta) \left[P_{j,j+1},\mathcal{O}\right]+\sin^2(\theta)\left(P_{j,j+1} \mathcal{O}P_{j,j+1}-\mathcal{O} \right)\,.
\end{align}
Suppose that the angles $\theta$ are randomly distributed such that $\overline{\sin^2(\theta)} = dt$ and $\overline{\sin(2\theta)} = 2J dt$, e.g. by setting $\overline{\theta^2} =dt$ and $\overline{\theta} = \sqrt{Jdt}$ in the limit of small $dt$, we find that the averaged operator dynamics satisfies
\begin{align}
d\overline{\mathcal{O}} = iJ dt \left[P_{j,j+1},\mathcal{O}\right]+dt \left(\mathcal{O}-P_{j,j+1} \mathcal{O}P_{j,j+1} \right)\,.
\end{align}
For a circuit acting on a full chain, we then recover the Lindblad equation from the main text.

\subsection{Self-energy}

Taking into account the antisymmetry of the two-operator component, the equations of motion \eqref{eq:1op} and \eqref{eq:2op} can be rewritten in terms of $S^{xy}_{m,n}  \equiv C^{xy}_{m,n}-C^{xy}_{n,m}$  as
\begin{align}
\partial_t C^z_n &= J \left[S^{xy}_{n-1,n}-S^{xy}_{n,n+1}\right] + \Delta_n C^z_n \,,\\
\partial_t S^{xy}_{m,n} & = 2 J \left[\delta_{m+1,n}\left(C^z_m-C^z_{m+1}\right)+\delta_{m,n+1}\left(C^z_m-C^z_{m-1}\right)\right] +\Delta_m S^{xy}_{m,n}+\Delta_n S^{xy}_{m,n}\,.
\end{align}
Performing a discrete Fourier transform, we can define
\begin{align}
C^z_{k} = \sum_{n} e^{i k n} C^z_n, \qquad S^{xy}_{k,q} = \sum_{m,n}  e^{i k m +i q n}S^{xy}_{m,n}\,.
\end{align}
where $k,q$ are momenta in the Brillouin zone, i.e. $k=2\pi n/N$ for $n = 0, \dots, N-1$. The resulting differential equations follow as
\begin{align}
\partial_t C^z_k &= -\Omega(k){C}^z_k - \frac{J}{N} \left[1-\exp\left(i k\right)\right] \sum_{q}\exp\left(i q\right) {S}^{xy}_{k+q,-q}\, ,\\
\partial_t S^{xy}_{k,q} &= -\left[ \Omega(k)+ \Omega(q) \right] {S}^{xy}_{k,q}+ 4 J \left[\cos\left(q\right)-\cos\left(k\right)\right]{C}^z_{k+q} \,.
\end{align}
where we have introduced 
\begin{equation}
\Omega(k) = 2 \left[1-\cos\left(k\right)\right] = 4 \sin^2\left(\frac{k}{2}\right)\,.
\end{equation}
Integrating the equation for $S^{xy}_{k-q,q} \times e^{[\Omega(q)+\Omega(k-q)]t}$ and plugging the resulting expression in the equation for $C^z_k$ returns
\begin{align}\label{eq:se-discrete}
  \partial_t C^z_k(t)  =  -\Omega(k){C}^z_k(t)-\frac{4J^2}{N}\sum_{q}\left[\cos(k-q)-\cos(q)\right]^2 \int_{0}^t ds \, e^{[\Omega(q)+\Omega(k-q)](s-t)} C^z_k(s)\,.
\end{align}
This is the time domain version of the Dyson equation $C^z(\omega,k)=\left[i\omega - \Omega(k)-\Sigma(\omega,k)\right]^{-1}$ with the self-energy $\Sigma(\omega,k)$ given by Eq.~\eqref{eq:se}. Note that this calculation is equivalent to a one-loop self-energy in the usual many-body perturbation theory language.

\subsection{Evaluating the self-energy}

After taking the continuum limit of Eq.~\eqref{eq:se}, we are left with the integral
\begin{equation}
  \Sigma(\omega,k) =  4J^2\int_{-\pi}^\pi \frac{dk_1}{2\pi} \frac{\left[\cos(k_1)-\cos(k-k_1)\right]^2}{\Omega(k_1)+\Omega(k-k_1)-i\omega}.
\end{equation}
Writing $z=e^{ik}$, $z_1=e^{ik_1}$ we can express this as the contour integral around the unit circle
\begin{equation}
  \Sigma(\omega,k) =  J^2\oint \frac{dz_1}{2\pi i} \frac{\left[z_1(1-z^{-1})+z_1^{-1}(1-z)\right]^2}{(4-i\omega)z_1 - z_1^2(1+z^{-1})-(1+z)}.
  \label{eq:contint}
\end{equation}
To evaluate this by the residue theorem, we identify 3 poles: one at $z_1=0$ due to the numerator and the other two due to the quadratic in the denominator. For the first we have
$$
\mathrm{Res}\left[\frac{\left[z_1(1-z^{-1})+z_1^{-1}(1-z)\right]^2}{(4-i\omega)z_1 - z_1^2(1+z^{-1})-(1+z)}\right]_{z_1=0} = (i\omega-4)\left(\frac{1-z}{1+z}\right)^2=(4-i\omega)\tan^2\frac{k}{2}.
$$
The roots of the quadratic are
$$
(4-i\omega)z_1 - z_1^2(1+z^{-1})-(1+z) = (z^{-1}+1)(z_1-z_+)(z_1-z_-)
$$
$$
z_\pm = \frac{\xi \pm \sqrt{\xi^2-16\cos^2k/2}}{2(1+e^{-ik})}
$$
where we've set $\xi\equiv 4-i\omega$. For $\xi$ real and in the range $-4\cos k/2<\xi < 4\cos k/2$ the square root is imaginary and the two poles lie on the unit circle $|z_\pm |=1$. This corresponds to a branch cut in the complex plane of $\omega$ on the imaginary axis. 

For $\mathrm{Im}\,\xi > 0$ $z_+$ is outside the unit circle and $z_-$ inside, with the situation reversing for $\mathrm{Im}\,\xi < 0$. Considering the former case, we have
$$
\mathrm{Res}\left[\frac{\left[z_1(1-z^{-1})+z_1^{-1}(1-z)\right]^2}{(4-i\omega)z_1 - z_1^2(1+z^{-1})-(1+z)}\right]_{z_1=z_-} = -\frac{\left[z_-(1-z^{-1})+z_-^{-1}(1-z)\right]^2}{\sqrt{\xi^2-16\cos^2k/2}}.
$$
Putting these two contributions together leads to the final result
\begin{equation}
  \Sigma(\omega,k) = J^2\left[(i\omega-4)\left(\frac{1-z}{1+z}\right)^2 -\frac{\left[z_-(1-z^{-1})+z_-^{-1}(1-z)\right]^2}{\sqrt{\xi^2-16\cos^2k/2}}\right]
  \label{eq:se-final}
\end{equation}
In the low momentum $(k\to 0)$ limit where $\omega$ is $O(k^2)$ the branch cut starts at $\omega \sim - ik^2/2$, $\sqrt{\xi^2-16\cos^2k/2}\sim 2\sqrt{k^2-2i\omega}$, and 
$$
z_-(1-z^{-1})+z_-^{-1}(1-z)\sim ik\sqrt{k^2-2i\omega}.
$$
After replacing $4-i\omega\to 4$ this gives Eq.~\eqref{eq:se-final-low}

\subsection{Diffusion constant}

The diffusion constant can be extracted from the Dyson equation Eq.~\eqref{eq:se-discrete} in the $N\to\infty$ limit,
\begin{align}\label{eq:int_diff_eq}
  \partial_t C^z(t,k)  =  -\Omega(k){C}^z(t,k)  - 4J^2 \int_{0}^{2\pi} \frac{dq}{2\pi}\left[\cos(k-q)-\cos(q)\right]^2 \int_{0}^t ds \, e^{\left[\Omega(k-q)+\Omega(q)\right](s-t)}C^z(s,k) \,.
\end{align}
Writing $C^z(t,k) = \tilde{C}^z(t,k) e^{-\Omega(k) t }$, we have that
\begin{align}\label{eq:diffusionconst_intpict}
D(t) = \left.-\frac{1}{2}  \frac{\partial}{\partial t} \frac{\partial^2}{\partial k^2}C^z(t,k) \right\vert_{k=0} = 1 - \frac{1}{2} \left. \frac{\partial}{\partial t} \frac{\partial^2}{\partial k^2}\tilde{C}^z(t,k) \right\vert_{k=0}\,.
\end{align}
Taking the second derivative of Eq.~\eqref{eq:int_diff_eq} w.r.t. $k$ then returns
\begin{align}\label{eq:int_diff_eq_2ndder}
\left. \frac{1}{2} \partial_{k}^2\partial_t \tilde{C}^z(t,k) \right\vert_{k=0} =  - 4 J^2 \int \frac{dq}{2\pi}\sin^2(k+q) \left. \int_{0}^t ds \,e^{-\left[\Omega(k)-\Omega(k+q)-\Omega(q) \right](s-t)}\tilde{C}^z(s,k) \right\vert_{k=0}\,,
\end{align}
where we have used that both $\left[\Omega(k+q)-\Omega(q)\right]^2$ and its first derivative vanish at $k=0$, such that the only nonvanishing contribution to the final expression is the one where the second derivative of this term is taken. Since $\partial_t \tilde{C}^z(t,k = 0) = 0$ and $\tilde{C}^z(t=0,k=0) = 1$, we find that $\tilde{C}^z(t,k=0) = 1$, such that for $k=0$  the integral over $s$ in Eq.~\eqref{eq:int_diff_eq_2ndder} can be explicitly evaluated to return 
\begin{align}
 \left.\frac{1}{2}\partial_{k}^2\partial_t  \tilde{C}^z(t,k) \right\vert_{k=0} =  - J^2 \int \frac{dq}{2\pi}\sin^2(q) \frac{1-e^{-4(1-\cos(q))t}}{1-\cos(q)} =-\frac{J^2}{2}  \oint \frac{dz}{2\pi i} \frac{(z+1)^2}{z^2}\left[1-e^{-4t + 2zt +2t/z}\right],
\end{align}
where in the second equality we have changed variables to $z = e^{iq}$ and the contour is the unit circle. The resulting contour integral can be evaluated using Cauchy's residue theorem, where there is a single pole at $z=0$. The exponential term is the generating function for the modified Bessel functions of the first kind, and we can make the residues at $z=0$ explicit by collecting the relevant orders
\begin{align}
e^{2zt +2t/z} &= \dots + I_1(4t) z + I_0(4t) + \frac{I_1(4t)}{z} + \dots,
\end{align}
such that contour integration return
\begin{align}
\left.\frac{1}{2} \partial_{k}^2\partial_t \tilde{C}^z(t,k) \right\vert_{k=0} &= -J^2 \left[1-e^{-4t}\left(I_0(4t)+I_1(4t)\right)\right]\,.
\end{align}
Combining this result with Eq.~\eqref{eq:diffusionconst_intpict} returns the time-dependent diffusion constant in Eq.~\eqref{eq:diffusionconstant}.

\end{widetext}

\end{document}